\documentclass[letterpaper,twocolumn,10pt]{article}
\usepackage{usenix,epsfig,endnotes}
\usepackage[table,xcdraw]{xcolor}
\usepackage[hyphens]{url}
\usepackage[blocks]{authblk}
\usepackage{amsmath}
\usepackage{amssymb}

\renewcommand{\paragraph}[1]{\vspace{0.1in}\noindent\textbf{#1}}

\begin{document}

\date{}

\title{\Large \bf Social Media COVID-19 Misinformation Interventions Viewed Positively, But Have Limited Impact}


\author[1]{Christine Geeng}
\author[1]{Tiona Francisco}
\author[2]{Jevin West}
\author[1]{Franziska Roesner}

\affil[1]{\small\it Paul G. Allen School of Computer Science \& Engineering, University of Washington, Seattle, WA, USA}
\affil[2]{\small\it Information School, University of Washington, Seattle, WA, USA}


\maketitle

\thispagestyle{empty}

\subsection*{Abstract}
Amidst COVID-19 misinformation spreading, social media platforms like Facebook and Twitter rolled out design interventions, including banners linking to authoritative resources and more specific “false information” labels. In late March 2020, shortly after these interventions began to appear, we conducted an exploratory mixed-methods survey ($N=311$) to learn: what are social media users’ attitudes towards these interventions, and to what extent do they self-report effectiveness? We found that most participants indicated a positive attitude towards interventions, particularly post-specific labels for misinformation. Still, the majority of participants discovered or corrected misinformation through other means, most commonly web searches, suggesting room for platforms to do more to stem the spread of COVID-19 misinformation.

\section{Introduction}

In late March 2020, social media platforms had recently increased implementation of misinformation interventions (such as banners or labels) in response to the proliferation of COVID-19 health misinformation.  Twitter and Facebook both added generic banners directing users to COVID-19 information, as well as added misinformation warnings to specific posts. To better understand user responses to these changes, we conducted a mixed-methods online survey, recruiting through Prolific and our personal networks, to gauge attitudes (of participants who had seen them) towards these interventions on Facebook, Instagram, and Twitter. Our survey was exploratory and no hypotheses were tested. We also collected accounts of how participants had learned that COVID-19 misinformation they had seen was false. Our research questions were:

\begin{enumerate}
    \item What are people’s attitudes towards social media platform interventions for COVID-19 misinformation, including generic banners linking to authoritative sources and specific false information labels? 
    \item How did people discover that COVID-19 misinformation was actually false? Specifically, what was the role of social media platform interventions in this discovery, compared to other methods?
\end{enumerate}

Our results show that participants rated the helpfulness of Facebook’s “False Information” label --- which appears on specific posts --- significantly higher than Facebook’s generic COVID-19 information banner, suggesting that post-specific interventions may be more effective. Some participants reacted negatively to the interventions, e.g., expressing a distrust of the platform. Despite the general acceptance of the interventions, we find 76.7\% of participants instead discovered information to be false through web searches or trusted health sites.  

Our results suggest that social media platform interventions are not yet doing the heaviest lifting when it comes to correcting misinformation, but people are receptive to these attempts. Our exploratory study raises open research questions, and our results suggest there is room for platforms to augment or support existing user strategies, as well as increase post-specific fact-check labeling.

\section{Related Work}

The rise of misinformation on social media has prompted sites like Facebook and Twitter to design platform affordances addressing misinformation, such as showing links to trusted public health sites for vaccine-related search terms~\cite{twitter-health}. Facebook has experimented with various interventions, ranging from showing “disputed” flags or “false information” labels on posts to more subtly showing fact-checking “related articles”~\cite{fb-disputed}. While having post-specific “disputed” labels might raise concerns over triggering the backfire effect~\cite{backfireEffect}, i.e., entrench existing false beliefs, recent replication~\cite{elusivebackfire} and review work suggests that ``backfire effects are not a robust empirical phenomenon''~\cite{SWIRETHOMPSON2020286}.
 
Findings about the effectiveness of interventions have been varied. Bode et al. found Facebook’s “related articles” to reduce health misperceptions~\cite{relatednews}. Pennycook et al. found that attaching warnings to fake news headlines could lead to incorrect belief that non-labeled headlines are not false~\cite{pennycook2020}. The latter study only displayed headlines to participants; we note that other work has shown that people use multiple heuristics on and off social media to determine information credibility~\cite{Flintham:2018,geeng2020,METZGER2013210}. Other approaches include pre-emptive debunking, which has been shown to be effective at preventing anti-vaccine conspiracy beliefs~\cite{Jolley2020}.

Since the COVID-19 pandemic began, Facebook, Twitter, and others have implemented more fact-checking affordances~\cite{twitter-covid,fb-covid}, given the various health misinformation that has arisen~\cite{snopes-breath}. To investigate the effectiveness of these interventions in this specific, currently highly-relevant context, we qualitatively and quantitatively surveyed sentiments of users who have seen these interventions on their own feeds, as well as their other experiences with COVID-19 misinformation. At the highest level, our results suggest that while most respondents are receptive to social media platforms’ attempts to curb COVID-19 misinformation, there remains room for improvement and future research to inform both platform designs and related policy discussions. 

\section{Methodology}

To answer our research questions, we conducted an anonymous online survey (approximately 10 minutes long) from March 20-26, 2020 to elicit quantitative and qualitative responses. Our study was reviewed and deemed exempt by the University of Washington Human Subjects Review Board (IRB). We did not collect identifying information about participants. For any quotes used in this paper, the quoted participant explicitly provided their consent (in the survey) to have their anonymized quotes used in publications.

To recruit participants, we used both Prolific, a paid crowdsourcing service, and our personal networks via social media. Prolific participants were paid \$13.86/hr, with an average 7 minute survey completion time. We also sought volunteers via our personal networks on Facebook and Twitter. Participants were screened out if they were not at least 18 years old or had not used Facebook, Twitter, or Instagram since March 1st (around the time when the COVID-19 misinformation interventions started to roll out). 

We recruited 111 participants through our personal networks and 202 through Prolific, and we removed 2 disingenuous responses (based on our review of answers to free-response questions), for a total of 311 completed surveys. In this paper, we discuss and analyze the results of both populations combined.

Demographic questions were optional. The majority of our participants were 18-24 years old (refer to Table~\ref{tab:age}). 37.94\% of our participants live in the United States, 10.29\% in Portugal, 9.97\% in the United Kingdom, 7.40\% in Canada, and 7.40\% in Poland. The rest of our participants live in a variety of other countries. Of the 118 participants living in the United States, 48 identify as Democrat, 11 as Republican, and 34 as Independent; the rest did not answer the question.

\begin{table}[t]
\centering
\begin{tabular}{cc}
\hline
\textbf{Age Range} & \textbf{Count} \\ \hline
18-24 & 130 \\
25-34 & 107 \\
35-44 & 36 \\
45-54 & 13 \\
55-64 & 13 \\
65-74 & 3 \\ \hline
\end{tabular}
\caption{Participant age ranges.}\label{tab:age}
\end{table}

\begin{figure}[t]
\centering
  \fbox{\includegraphics[width=.9\columnwidth]{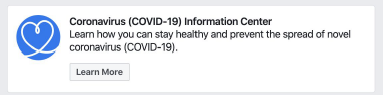}}\vspace{0.2in}
  \fbox{\includegraphics[width=.9\columnwidth]{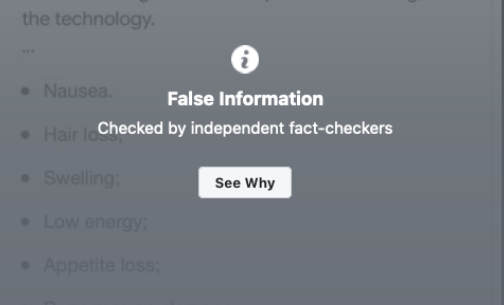}}
  \caption{Facebook’s design interventions. On top is a generic banner shown when someone searches for COVID-19 or related terms, and on the bottom is a label for posts labeled as false by Facebook’s fact checking partners.}\label{fig:facebook}
\end{figure}

\begin{figure}[t]
\centering
  \fbox{\includegraphics[width=.9\columnwidth]{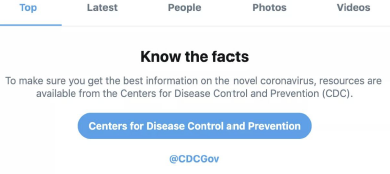}}\vspace{0.2in}
  \fbox{\includegraphics[width=.9\columnwidth]{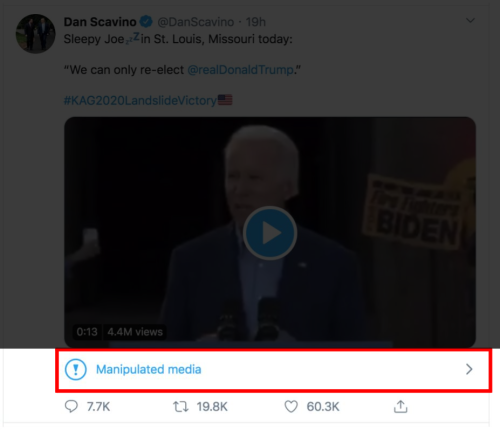}}
  \caption{Twitter’s design interventions. On top is a generic banner shown when someone searches for COVID-19 or related terms, and on the bottom is a label for posts known to Twitter to contain manipulated media.}\label{fig:twitter}
\end{figure}

\begin{figure}[t]
\centering
  \fbox{\includegraphics[width=.9\columnwidth]{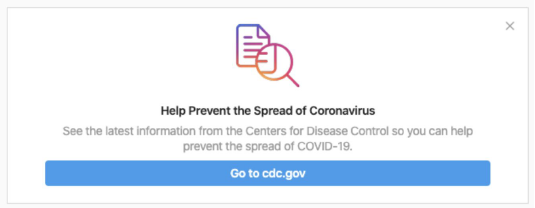}}
  \caption{Instagram’s design intervention: a generic banner shown when someone searches for COVID-19 or related terms.}\label{fig:instagram}
\end{figure}

Our survey asked if participants had seen Facebook, Twitter, or Instagram COVID-19 or misinformation interventions (circa March 2020) before. If so, we asked both an open-ended question about their thoughts, as well as a question about how helpful they considered the intervention, on a 5-point scale from “Not at all helpful” (1) to “Extremely helpful” (5). For interventions that labeled specific misinformation, we asked how that had changed the participant’s view of the post, if at all. The screenshots we showed participants of the interventions we asked about are shown in Figures~\ref{fig:facebook}-\ref{fig:instagram}.

We also asked for anecdotes of when participants had seen or believed COVID-19 misinformation, where they had seen it, how they discovered its falsity, and what they did upon realizing this. Finally,  we asked participants to select from a list of known COVID-19 misinformation which they had seen. 

To analyze open-ended responses about perceptions of interventions, three coders independently inductively coded a subset of these answers before discussing and agreeing on a codebook of 17 codes. Following McDonald et al.’s guidelines on when to seek coding agreement~\cite{mcdonald2019reliability}, we double coded a subset (46.34\% of total responses) to check for agreement and then had a single coder code the rest of the responses. For the double-coded subset, we calculated Cohen's $\kappa$ for inter-coder reliability, given that we had two coders and nominal data~\cite{kappa}. We had a $\kappa$ of ``substantial'' (0.61–0.80) to ``almost perfect agreement'' (0.81–1.00) for 87.5\% of categories (see Appendix). We discussed code usage discrepancies between coders until we reached a consensus in the final coding. 

To analyze our helpfulness scale data, we compared helpfulness ratings between interventions from the same site by comparing between participants who had seen both. Since our scale data is ordinal, we used a Wilcoxon signed rank test and only report on tests with significance.

\section{Results}
Our results reveal a variety of reactions to misinformation labels and different modes of discovering misinformation.

\paragraph{Social media interventions are used, but are outweighed by other strategies for debunking misinformation.}
When asked in a multiple-response question how they learned something they saw was false (whether or not they initially believed it), participants told us most frequently that they conducted a web search (39.6\% of 240 who answered this question), sought out trusted sources (37.1\%), saw a correction in a social media comment (19.2\%), or heard a correction from someone directly (12.1\%). Only 4.2\% learned something was false because the social media platform had labeled it as such. The majority (71.7\%) of respondents indicated they ``knew it wasn't true'', though we cannot verify whether respondents’ baseline knowledge was correct. 

\paragraph{Post-specific social media interventions are viewed as more helpful, and seem to be more effective, than generic interventions for our participants.} We find that participants tended to rate (on a 1-to-5 scale) post-specific interventions as more helpful. For example, comparing the 30 participants who had seen both Facebook interventions, these participants found the post-specific “False Information” label significantly more helpful (median rating of 4 “very helpful”) than the generic banner (median rating of 2 “slightly helpful”). (Wilcoxon signed-rank test, $V$ = 4, $Z$ = -4.13, $p$ = 0.018, $r$ = 0.75).

Considering effectiveness, only 13.3\% of 105 participants who saw the Facebook banner said that they had ever clicked on it. Meanwhile, 32.3\% of the 65 participants who saw the Facebook “False Information” label said they no longer believe the content of the post due to the label. 50.8\% self-reported (albeit in retrospect) that they had already not believed the false-labeled post, and only 6.2\% said that they continued to believe the post, or believed it more, given the label.

The median helpfulness rating of the generic Twitter banner was 3 (“somewhat helpful”), with a reported clickthrough rate of 32.8\% by 58 participants who saw the banner. The difference in helpfulness rating between the Facebook and Twitter banners (medians of 2 and 3 respectively), for the 26 participants who saw both, was not statistically significant (Wilcoxon signed-rank test, $V$ = 4.5, $Z$ = -2.11, $p$ = 0.06, $r$ = 0.41). 

\begin{table}[tb]
\begin{tabular}{lr}
\hline
\textbf{Reactions} & \textbf{Count} \\ \hline
\rowcolor[HTML]{EFEFEF} 
Positive & 150 \\
Nothing/Neutral & 39 \\
\rowcolor[HTML]{EFEFEF} 
Unnecessary because I'm already informed & 27 \\
I ignored it & 19 \\
\rowcolor[HTML]{EFEFEF} 
Other & 16 \\
Annoying & 12 \\
\rowcolor[HTML]{EFEFEF} 
Fear/worry about the future & 11 \\
It's common to see these banners now & 10 \\
\rowcolor[HTML]{EFEFEF} 
Cautious/suspicious of the banner & 8 \\
\begin{tabular}[c]{@{}l@{}}I don't trust the company, so I do not trust \\ the banner\end{tabular} & 7 \\
\rowcolor[HTML]{EFEFEF} 
Done too late & 6 \\
It looked official & 6 \\
\rowcolor[HTML]{EFEFEF} 
Thought it was a ad & 4 \\
\begin{tabular}[c]{@{}l@{}}Don't know enough to comment on \\ execution\end{tabular} & 4 \\
\rowcolor[HTML]{EFEFEF} 
Angry & 3 \\
Could be abused to censor & 2 \\ \hline
\end{tabular}
\caption{Counts of qualitatively coded reactions to ``What did this intervention make you think or feel?'' for all social media interventions we showed participants, out of 246 responses.}\label{tab:thinkgorfeel}
\end{table}

\paragraph{Most participants had positive or neutral responses to social media platform misinformation interventions.} We collected qualitative free-response data about participants’ attitudes and highlight key themes here. We find participants’ opinions about platform interventions range from positive (“I thought it was good that Facebook was trying to do something to inform people better”) to neutral (“I didn't think much of it. I follow the news so I didn't click on this one because I already know the basic details”) to negative (“I don't like it. I don't need Facebook to tell me this, and I don't trust their automated way of detecting it”\footnote{We note that as of at least June 2020, on Facebook, misinformation is labeled as false only after a human fact-checker reviews it~\cite{fb-factchecking}.}) to --- rarely --- hostile (“I was irritated because it is another in a long list of ‘tools’ to ‘protect’ users. In my opinion, this label assumes people are morons and unable to discern what’s true, false and/or misleading”). 

Table \ref{tab:thinkgorfeel} shows how often themes we coded appeared in responses. Our analysis focused on negative reactions, as these provide more actionable information.  A sentiment expressed by both positively and negatively-reacting participants was that they found the interventions unnecessary, because they were already sufficiently informed about COVID-19.

\begin{table}[tb]
\begin{tabular}{lrr}
\hline
\textbf{Choice} & \textbf{\%} & \textbf{Count} \\ \hline
\rowcolor[HTML]{EFEFEF} 
Nothing & 54.62\% & 130 \\
Corrected the person publicly & 18.49\% & 44 \\
\rowcolor[HTML]{EFEFEF} 
Corrected the person privately & 16.81\% & 40 \\
Other & 11.34\% & 27 \\
\rowcolor[HTML]{EFEFEF} 
I don't know/remember & 3.36\% & 8 \\ \hline
\end{tabular}
\caption{Multiple-choice responses to “What did you do when you realized COVID-19 information someone else shared was false?” $N$ = 238.}\label{tab:someoneelse}
\end{table}

\begin{table}[tb]
\begin{tabular}{lrr}
\hline
\textbf{Choice} & \textbf{\%} & \textbf{Count} \\ \hline
\rowcolor[HTML]{EFEFEF} 
Nothing & 57.28\% & 59 \\
Shared the correction & 27.18\% & 28 \\
\rowcolor[HTML]{EFEFEF} 
Other & 12.62\% & 13 \\
I don’t know/remember & 8.74\% & 9 \\
\rowcolor[HTML]{EFEFEF} 
Unshared it if you had shared it & 1.94\% & 2 \\ \hline
\end{tabular}
\caption{Multiple-choice responses to “What did you do when you realized COVID-19 information that you believed was false?” $N$ = 103.}\label{tab:youbelieved}
\end{table}

\paragraph{When participants came across misinformation and realized it was false, 54.62\% did nothing, but 35.3\% made a correction.} Our results suggest that COVID-19 misinformation was rampant on social media and the web in late March, 2020: 79.5\% of participants reported having seen others share COVID-19 related misinformation, and 33.9\% reported believing something false themselves. Table~\ref{tab:someoneelse} and Table~\ref{tab:youbelieved} show participants’ self-reported reactions when they realized they or their contacts had shared COVID-19 misinformation. In both cases, a slight majority of participants did nothing, though a significant fraction also publicly or privately shared a correction.   

Public corrections sometimes occurred in group chats. One participant stated, “On the same group where the message was shared, with my friends, we discussed the fact that it was false after it appeared on the news.” Others added comments with corrections to posts or “liked” an existing correction. Private corrections involved in-person conversations, email, or direct messages. Some “Other” responses included reporting the post or filtering unwanted content from one’s social media feed.

Though we did not collect data on reasons for taking no action, we note that these reasons might include not wanting to engage in a debate, not being able to find the original post again, not considering the issue personally relevant enough, or not having re-shared the false information themselves after believing it.

\section{Discussion}
From our results, we make some suggestions towards improving misinformation labeling efforts.

\paragraph{Social media platforms should increase specific misinformation labeling efforts.} In the context of COVID-19, our participants had generally positive responses to the interventions, and found specific misinformation labels to be more helpful than generic banners pointing to authoritative sources. We also found that these specific labels worked for many participants: out of the 65 people who saw the Facebook label, 21 people heeded the label, while only 2 people continued to believe the post and 2 people believed it more. Our results thus suggest that --- at least among our study population --- the labels generally produce the intended effect rather than a “backfire effect”~\cite{backfireEffect,correctingmisinfo}; this supports other work that find no robust evidence for this phenomenon~\cite{SWIRETHOMPSON2020286,elusivebackfire}. However, only 4.2\% of respondents who said they have seen misinformation stated they learned it was not true through social media labeling, suggesting that they often see misinformation on social media that is not labeled by the platform. This finding suggests a strong motivation for social media platforms to significantly increase the amount and frequency of misinformation that they explicitly label as false.

\paragraph{Authoritative banners should be designed to not look like ads, and warning fatigue should be considered.} While the banners were the result of collaborations between social media sites and the World Health Organization and other national public health agencies~\cite{twitter-covid,fb-covid}, 11 responses (out of 246 responses) mentioned the banners looked like ads or that they did not trust the social media company enough to trust the banner. The sheer frequency with which participants see these or similar banners across different sites may also lead to warning fatigue~\cite{bravo-lillo}. Indeed, 27 responses noted ignoring the banner because they have already seen so much other information about COVID-19. Future research should further study these effects and how to avoid them.

\paragraph{Open research questions remain around intervention design and effectiveness, side effects, and interventions beyond COVID-19.} As discussions around platform responsibility and potential liability in the face of misinformation intensify, policymakers will need substantive evidence to lean on to inform these discussions. Our results suggest that different interventions have different impacts, so multiple and continued studies are needed. We call on future research to help answer these open questions, considering different types of users, different types of content, and different types of intervention designs.

For example, our study does not attempt to differentiate different types of people, who may react differently to interventions. A Pew research study showed that Americans engage with online information in varying ways, ranging from eager or curious to distrustful of information sources~\cite{Pew:approachfacts}. Future research should study whether interventions like the ones we study here are most effective for certain types of information consumers --- for example, people who trust the fact-checking sources used by social media platforms, and people who are not already convinced of the relevant misinformation but are attempting to become informed. Different intervention designs may be effective for different information consumers.  

Future work should also explore if there are other potential side effects of the interventions, beyond debunking misinformation directly.  For example, while the generic banners may not change behaviors in the moment, perhaps they have a more subtle, sustained impact on how people evaluate information in their feeds. This impact might be positive (reducing trust in misinformation) but may also be negative, e.g., increasing trust in misinformation that doesn’t have a fact-checking label~\cite{pennycook2020}.  

We found that 35.3\% of participants corrected others sharing misinformation, and 27.18\% of participants shared a correction when they themselves had believed misinformation. These numbers are far below 100\%, but non-trivial. Future research and design should explore how much room there is to increase (self-)correcting behavior from users, experimenting with ways to make sharing corrections easier.

Finally, COVID-19 is a unique situation, and future research should study how people use and react to platform-based interventions on other topics (e.g., political misinformation, climate change). People may consider certain platform-based interventions more appropriate during a global pandemic, but may prefer that social media platforms take a less active role in labeling content in other circumstances. The normalization of current platform practices may also shift user perspectives for the future.

\section{Limitations}
Our exploratory study is based on a convenience sample of participants and our results may not be generalizable to broader, more representative populations. Most of our participants live in the United States; individuals living in other countries may have seen other misinformation more relevant to their geographic location that we did not ask about, or different versions of the platform interventions than the screenshots we showed. For participants sampled from our personal networks, they may have skewed towards academics and people with an interest in computer security and privacy. With any self-report methodology, responses are susceptible to recall bias, influence from wording, and erroneous statements~\cite{paulhus2007self}. We did not compare differences in data between our two sampling populations as it is unclear what variations and similarities there are between these two groups. We make no strong quantitative claims about our qualitative results. Finally, the COVID-19 situation and platform interventions themselves are changing rapidly; our results represent one snapshot in time (late March 2020). Nevertheless, this study sheds light on participants’ reactions to platform interventions in a hotly-debated and quickly evolving space, and raises new research questions and directions for future work.
\section{Conclusion}
To better understand people's responses to social media platform interventions for COVID-19 misinformation, we conducted an exploratory mixed-methods online survey in late March 2020 to gauge attitudes (of participants who had seen them) towards these interventions on Facebook, Instagram, and Twitter, as well as to collect accounts of how participants have learned COVID-19 misinformation was false.
Our results suggest that post-specific interventions may be more effective. and that social media platform interventions are not yet doing the heaviest lifting when it comes to correcting misinformation, but people are receptive to these attempts.

\section{Acknowledgements}
We thank Tadayoshi Kohno, Lucy Simko, and Miranda Wei for their feedback on our survey instrument, and Yim Register for feedback on an earlier version of this paper. This paper was supported in part by the National Science Foundation under Award CNS-1651230 and the John S. and James L. Knight Foundation.

{\footnotesize \bibliographystyle{acm}
\bibliography{main}}

\begin{thebibliography}{10}

\bibitem{relatednews}
{\sc Bode, L., and Vraga, E.~K.}
\newblock In related news, that was wrong: The correction of misinformation
  through related stories functionality in social media.
\newblock {\em Journal of Communication 65}, 4 (2015), 619--638.

\bibitem{bravo-lillo}
{\sc Bravo-Lillo, C., Cranor, L., Komanduri, S., Schechter, S., and Sleeper,
  M.}
\newblock Harder to ignore? revisiting pop-up fatigue and approaches to prevent
  it.
\newblock In {\em 10th Symposium On Usable Privacy and Security ({SOUPS}
  2014)\/} (Menlo Park, CA, July 2014), {USENIX} Association, pp.~105--111.

\bibitem{fb-disputed}
{\sc {Facebook}}.
\newblock {Replacing Disputed Flags With Related Articles}, Dec. 2017.
\newblock
  \url{https://newsroom.fb.com/news/2017/12/news-feed-fyi-updates-in-our-fight-against-misinformation/}.

\bibitem{fb-factchecking}
{\sc Facebook}.
\newblock Facebook’s approach to fact-checking: How it works, Aug. 2020.
\newblock
  \url{https://www.facebook.com/journalismproject/programs/third-party-fact-checking/how-it-works}.

\bibitem{Flintham:2018}
{\sc Flintham, M., Karner, C., Bachour, K., Creswick, H., Gupta, N., and Moran,
  S.}
\newblock Falling for fake news: Investigating the consumption of news via
  social media.
\newblock In {\em Proceedings of the 2018 CHI Conference on Human Factors in
  Computing Systems\/} (New York, NY, USA, 2018), CHI '18, ACM,
  pp.~376:1--376:10.

\bibitem{geeng2020}
{\sc Geeng, C., Yee, S., and Roesner, F.}
\newblock Fake news on facebook and twitter: Investigating how people (don't)
  investigate.
\newblock In {\em Proceedings of the 2020 CHI Conference on Human Factors in
  Computing Systems\/} (New York, NY, USA, 2020), CHI '20, Association for
  Computing Machinery, p.~1–14.

\bibitem{Pew:approachfacts}
{\sc Horrigan, J.~B.}
\newblock How people approach facts and information, 2017.
\newblock
  \url{https://www.pewresearch.org/internet/2017/09/11/how-people-approach-facts-and-information/}.

\bibitem{Jolley2020}
{\sc Jolley, D., and Douglas, K.~M.}
\newblock Prevention is better than cure: Addressing anti-vaccine conspiracy
  theories.
\newblock {\em Journal of Applied Social Psychology 47}, 8 (2017), 459--469.

\bibitem{snopes-breath}
{\sc Kasprak, A.}
\newblock No, holding your breath is not a 'simple self-check' for coronavirus,
  Aug. 2018.
\newblock \url{https://www.snopes.com/fact-check/taiwan-experts-self-check/}.

\bibitem{correctingmisinfo}
{\sc Lewandowsky, S., Ecker, U. K.~H., Seifert, C.~M., Schwarz, N., and Cook,
  J.}
\newblock Misinformation and its correction: Continued influence and successful
  debiasing.
\newblock {\em Psychological Science in the Public Interest 13}, 3 (2012),
  106--131.
\newblock PMID: 26173286.

\bibitem{mcdonald2019reliability}
{\sc McDonald, N., Schoenebeck, S., and Forte, A.}
\newblock Reliability and inter-rater reliability in qualitative research:
  Norms and guidelines for cscw and hci practice.
\newblock {\em Proceedings of the ACM on Human-Computer Interaction 3}, CSCW
  (2019), 1--23.

\bibitem{kappa}
{\sc McHugh, M.~L.}
\newblock Interrater reliability: the kappa statistic.
\newblock {\em Biochemia Medica 22}, 3 (Oct. 2012).

\bibitem{METZGER2013210}
{\sc Metzger, M.~J., and Flanagin, A.~J.}
\newblock Credibility and trust of information in online environments: The use
  of cognitive heuristics.
\newblock {\em Journal of Pragmatics 59\/} (2013), 210 -- 220.
\newblock Biases and constraints in communication: Argumentation, persuasion
  and manipulation.

\bibitem{backfireEffect}
{\sc Nyhan, B., and Reifler, J.}
\newblock When corrections fail: The persistence of political misperceptions.
\newblock {\em Political Behavior 32\/} (June 2010), 303--330.

\bibitem{paulhus2007self}
{\sc Paulhus, D.~L., and Vazire, S.}
\newblock The self-report method.
\newblock {\em Handbook of research methods in personality psychology 1\/}
  (2007), 224--239.

\bibitem{pennycook2020}
{\sc Pennycook, G., Bear, A., Collins, E.~T., and Rand, D.~G.}
\newblock The implied truth effect: Attaching warnings to a subset of fake news
  headlines increases perceived accuracy of headlines without warnings.
\newblock {\em Management Science 66}, 11 (2020), 4944--4957.

\bibitem{twitter-covid}
{\sc Policy, T.~P.}
\newblock Stepping up our work to protect the public conversation around
  covid-19, 2020.
\newblock
  \url{https://blog.twitter.com/en_us/topics/company/2020/stepping-up-our-work-to-protect-the-public-conversation-around-covid-19.html}.

\bibitem{fb-covid}
{\sc Rosen, G.}
\newblock An update on our work to keep people informed and limit
  misinformation about covid-19, 2020.
\newblock \url{https://about.fb.com/news/2020/04/covid-19-misinfo-update/}.

\bibitem{SWIRETHOMPSON2020286}
{\sc Swire-Thompson, B., DeGutis, J., and Lazer, D.}
\newblock Searching for the backfire effect: Measurement and design
  considerations.
\newblock {\em Journal of Applied Research in Memory and Cognition" 9}, 3
  (2020), 286 -- 299.

\bibitem{twitter-health}
{\sc {Twitter}}.
\newblock {Helping you find reliable public health information on Twitter}, May
  2019.
\newblock
  \url{https://blog.twitter.com/en_us/topics/company/2019/helping-you-find-reliable-public-health-information-on-twitter.html}.

\bibitem{elusivebackfire}
{\sc Wood, T., and Porter, E.}
\newblock The elusive backfire effect: Mass attitudes' steadfast factual
  adherence.
\newblock {\em Political Behavior 41\/} (01 2018).

\end{thebibliography}

\appendix
\section*{Appendix}\label{Appendix}

\section{Survey Instrument}\label{appendix:survey}
\begin{enumerate}
\item When did you last use each social media site? 

After March 1st	Before March 1st	Never

Facebook 	$\circ$ 	$\circ$ 	$\circ$ 
Twitter 	$\circ$ 	$\circ$ 	$\circ$ 
Instagram 	$\circ$ 	$\circ$ 	$\circ$ 

\vspace{0.1in}
\begin{figure}[h]
\centering
  \fbox{\includegraphics[width=.9\columnwidth]{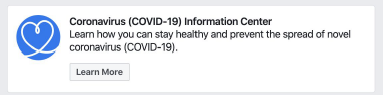}}
\end{figure}\vspace{-0.1in}

\item Have you seen this banner on Facebook before?
 
$\circ$ Yes 
$\circ$ No 
$\circ$ I don't know

\item You said you've seen this banner on Facebook before. What did you feel or think about it?

\item How helpful was this banner?
$\circ$ Extremely helpful 
$\circ$ Very helpful 
$\circ$ Somewhat helpful 
$\circ$ Slightly helpful 
$\circ$ Not at all helpful

\item Did you click on the banner?
$\circ$ Yes 
$\circ$ No 
$\circ$ I don't know/remember 

\vspace{0.7in}

\begin{figure}[h]
\centering
  \fbox{\includegraphics[width=.9\columnwidth]{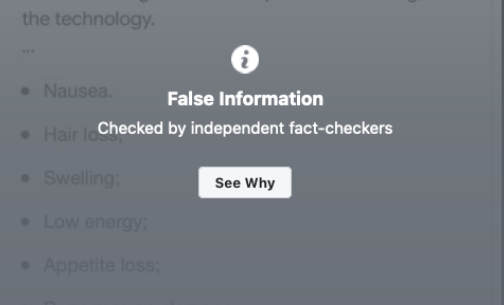}}\vspace{-0.1in}
\end{figure}

\item Have you seen this "False Information" label on Facebook before?
 
$\circ$ Yes 
$\circ$ No 
$\circ$ I don't know 

\item You said you've seen this "False Information" label on Facebook before. What did you think or feel about it?

\item How helpful was this label?
$\circ$ Extremely helpful 
$\circ$ Very helpful 
$\circ$ Somewhat helpful 
$\circ$ Slightly helpful 
$\circ$ Not at all helpful 

\item Think of a recent time when you saw this label. Did the label change your view of the post it was referring to? 
$\circ$ Yes, I no longer believed the post 
$\circ$ Yes, I believed the post more 
$\circ$ No, I already didn't believe the post 
$\circ$ No, I still believe the post 
$\circ$ Other 

\vspace{0.1in}
\begin{figure}[h]
\centering
  \fbox{\includegraphics[width=.9\columnwidth]{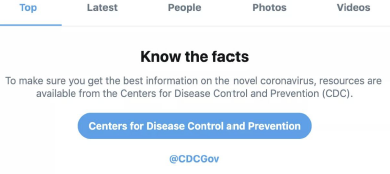}}
\end{figure}\vspace{-0.1in}

\item Have you seen this banner on Twitter before?
 
$\circ$ Yes 
$\circ$ No 
$\circ$ I don't know 

\item You said you've seen this banner on Twitter before. What did you think or feel about it?

\item How helpful was this banner?
$\circ$ Extremely helpful 
$\circ$ Very helpful 
$\circ$ Somewhat helpful 
$\circ$ Slightly helpful 
$\circ$ Not at all helpful 

\item Did you click on this banner?
$\circ$ Yes  
$\circ$ No 
$\circ$ I don't know/remember 

\vspace{0.5in}

\begin{figure}[h]
\centering
  \fbox{\includegraphics[width=.9\columnwidth]{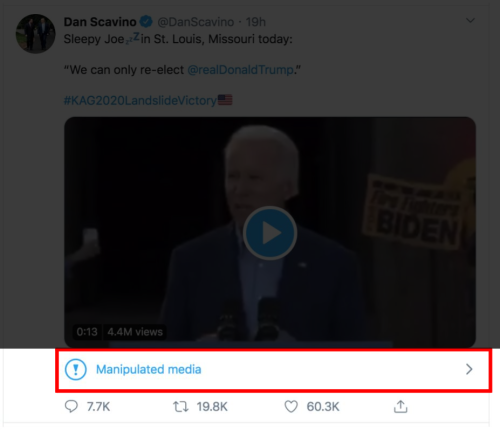}}\vspace{-0.1in}
\end{figure}

\item Have you seen this "Manipulated media" label on Twitter before?
  
$\circ$ Yes 
$\circ$ No 
$\circ$ I don't know 

\item You said you've seen this "Manipulated media" label on Twitter before. What did you think or feel about it? 
 
\item How helpful was this label?
$\circ$ Extremely helpful 
$\circ$ Very helpful 
$\circ$ Somewhat helpful 
$\circ$ Slightly helpful 
$\circ$ Not at all helpful 

\item Think of a recent time when you saw this label. Did the label change your view of the post it was referring to? 
$\circ$ Yes, I no longer believed the post 
$\circ$ Yes, I believed the post more 
$\circ$ No, I already didn't believe the post 
$\circ$ No, I still believe the post 
$\circ$ Other 

\vspace{0.1in}
\begin{figure}[h]
\centering
  \fbox{\includegraphics[width=.9\columnwidth]{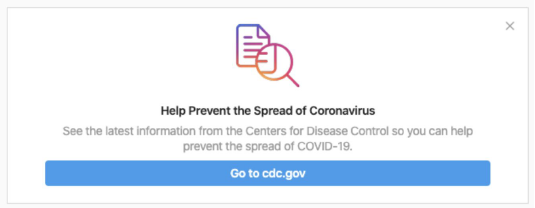}}
\end{figure}\vspace{-0.1in}

\item Have you seen this banner on Instagram before?
 
$\circ$ Yes 
$\circ$ No 
$\circ$ I don't know 

\item You mentioned that you've seen this banner on Instagram before. What did you think or feel about it?

\item How helpful was this banner?
$\circ$ Extremely helpful 
$\circ$ Very helpful 
$\circ$ Somewhat helpful 
$\circ$ Slightly helpful 
$\circ$ Not at all helpful 

\item Did you click on this banner?
$\circ$ Yes 
$\circ$ No 
$\circ$ I don't know/remember 

\item To the best of your knowledge, have you believed any misinformation about COVID-19?
$\circ$ Yes 
$\circ$ No 

For the following questions, think of one instance where you had believed misinformation about COVID-19.

\item Where did you see/hear this misinformation?
$\circ$ Social media site (e.g. Facebook, TikTok, etc.) 
$\circ$ News site 
$\circ$ Messaging app/group chat 
$\circ$ Word of mouth 
$\circ$ I don't know/remember 
$\circ$ Other 

\item How did you discover this information was false? (Select all that apply.)
$\Box$	I knew it wasn't true 
$\Box$	The website/platform itself labeled it as misinformation 
$\Box$	I conducted a web search (e.g. checking for multiple sources that agree or a fact-checking site) 
$\Box$	I looked at a trusted health site (such as WHO.int or CDC.gov) 
$\Box$	I saw a social media comment where someone debunked it 
$\Box$	Someone told me directly it was false (in-person, personal chat, etc.) 
$\Box$	I don't know/remember 
$\Box$	Other 

\item What was the misinformation?

\item What did you do when you realized this information was false?
$\Box$	Nothing 
$\Box$	Unshared it if you had shared it (please elaborate): 
$\Box$	Shared the correction (please elaborate): 
$\Box$	Other (please elaborate): 
$\Box$	I don’t know/remember 

\item To the best of your knowledge, have you noticed others sharing misinformation about COVID-19?
$\circ$ Yes 
$\circ$ No

\item For the following questions, think of one instance where you had noticed others sharing misinformation about COVID-19.

\item Where did you see/hear this misinformation?
$\circ$ Social media site (e.g. Facebook, TikTok, etc.) 
$\circ$ News site 
$\circ$ Messaging app/group chat 
$\circ$ Word of mouth 
$\circ$ I don't know/remember 
$\circ$ Other 

\item How did you discover this information was false? (Select all that apply.)
$\Box$	I knew it wasn't true 
$\Box$	The website/platform itself labeled it as misinformation 
$\Box$	I conducted a web search (e.g. checking for multiple sources that agree or a fact-checking site) 
$\Box$	I looked at a trusted health site (such as WHO.int or CDC.gov) 
$\Box$	I saw a social media comment where someone debunked it 
$\Box$	Someone told me directly it was false (in-person, personal chat, etc.) 
$\Box$	I don't know/remember 
$\Box$	Other 

\item What was the misinformation?

\item What did you do when you realized this information was false?
$\Box$	Nothing 
$\Box$	Corrected the person privately (please elaborate): 
$\Box$	Corrected the person publicly (please elaborate): 
$\Box$	Other (please elaborate): 
$\Box$	I don't know/remember 

\item Please check below all of the COVID-19 rumors that you have heard (whether or not you thought they might be true).

(Please be aware that these are all false rumors. For up-to-date information on the virus, please go to WHO.int or CDC.gov.)
$\Box$	The novel coronavirus sickness is caused by 5G 
$\Box$	There’s a plot to “exterminate” people infected with the new coronavirus 
$\Box$	Scientists have proven that humans got the novel coronavirus from eating bats 
$\Box$	Scientists predicted the virus will kill 65 million people 
$\Box$	China built a biological weapon that was leaked from a lab in Wuhan 
$\Box$	Chinese spies smuggled the virus out of Canada 
$\Box$	A coronavirus vaccine already exists 
$\Box$	There were 100,000 confirmed cases in January 
$\Box$	A teen on TikTok is the first case in Canada 
$\Box$	There will be a mass quarantine and martial law in a certain state (e.g. Washington) 
$\Box$	Other 

\item Can we use anonymized quotes from your free-response answers in future research publications?
$\circ$ Yes 
$\circ$ No 

\item In which country do you currently reside?
$\circ$ United States of America ... Zimbabwe

\item In which state do you currently reside?
$\circ$ Alabama ... I do not reside in the United States

\item What is your political affiliation?
$\circ$ Democrat 
$\circ$ Republican 
$\circ$ Independent 
$\circ$ Other
$\circ$ Not applicable

\item What gender(s) do you identify as?
$\Box$	Male 
$\Box$	Female 
$\Box$	Non-binary 
$\Box$	Prefer to self-describe:

\item What is your age?
$\circ$ 18-24 years old 
$\circ$ 25-34 years old 
$\circ$ 35-44 years old 
$\circ$ 45-54 years old 
$\circ$ 55-64 years old 
$\circ$ 65-74 years old 
$\circ$ 75 years or older

\item Anything you want to tell us? (Not required.)

\end{enumerate}

\onecolumn

\section{Inter-Rater Reliability}\label{appendix:irr}
Table~\ref{tab:coding} shows the inter-rater reliability percentages for our qualitative codes.
\vspace{0.3in}

\begin{table*}[h!]
\centering
\begin{tabular}{lrr}
\hline
{\color[HTML]{000000} Code} & {\color[HTML]{000000} Percent Agreement} & {\color[HTML]{000000} Cohen's Kappa} \\ \hline
\rowcolor[HTML]{EFEFEF} 
{\color[HTML]{000000} Nothing/Neutral} & {\color[HTML]{000000} 96.52\%} & {\color[HTML]{000000} 83.73\%} \\
{\color[HTML]{000000} Positive} & {\color[HTML]{000000} 100\%} & {\color[HTML]{000000} undefined*} \\
\rowcolor[HTML]{EFEFEF} 
{\color[HTML]{000000} Fear/worry about the future (generic)} & {\color[HTML]{000000} 99.13\%} & {\color[HTML]{000000} 90.46\%} \\
{\color[HTML]{000000} It's common to see these banners now} & {\color[HTML]{000000} 100\%} & {\color[HTML]{000000} 100.00\%} \\
\rowcolor[HTML]{EFEFEF} 
{\color[HTML]{000000} Thought it was a ad} & {\color[HTML]{000000} 98.26\%} & {\color[HTML]{000000} 65.77\%} \\
{\color[HTML]{000000} Unnecessary because I'm already informed} & {\color[HTML]{000000} 98.26\%} & {\color[HTML]{000000} 86.58\%} \\
\rowcolor[HTML]{EFEFEF} 
{\color[HTML]{000000} Annoying} & {\color[HTML]{000000} 99.13\%} & {\color[HTML]{000000} 88.44\%} \\
{\color[HTML]{000000} Done too late} & {\color[HTML]{000000} 97.39\%} & {\color[HTML]{000000} 65.33\%} \\
\rowcolor[HTML]{EFEFEF} 
{\color[HTML]{000000} Cautious/suspicious of the banner} & {\color[HTML]{000000} 95.65\%} & {\color[HTML]{000000} 42.44\%} \\
{\color[HTML]{000000} I don't trust the company, so I do not trust the banner} & {\color[HTML]{000000} 99.13\%} & {\color[HTML]{000000} 79.57\%} \\
\rowcolor[HTML]{EFEFEF} 
{\color[HTML]{000000} It looked official} & {\color[HTML]{000000} 99.13\%} & {\color[HTML]{000000} 66.28\%} \\
{\color[HTML]{000000} Angry} & {\color[HTML]{000000} 100\%} & {\color[HTML]{000000} 100.00\%} \\
\rowcolor[HTML]{EFEFEF} 
{\color[HTML]{000000} Don't know enough to comment on execution} & {\color[HTML]{000000} 100\%} & {\color[HTML]{000000} 100.00\%} \\
{\color[HTML]{000000} Could be abused to censor} & {\color[HTML]{000000} 100\%} & {\color[HTML]{000000} 100.00\%} \\
\rowcolor[HTML]{EFEFEF} 
{\color[HTML]{000000} I ignored it} & {\color[HTML]{000000} 98.26\%} & {\color[HTML]{000000} 49.12\%} \\
{\color[HTML]{000000} Other} & {\color[HTML]{000000} 89.57\%} & {\color[HTML]{000000} 8.73\%} \\ \hline
\end{tabular}
\caption{\centering Inter-rater reliability percentages.
\protect\url{*http://dfreelon.org/2008/10/24/recal-error-log-entry-1-invariant-values/} \label{tab:coding}}
\end{table*}


\end{document}